\documentclass[10pt]{iopart}
\usepackage{graphicx}   %       for graphics
\usepackage{latexsym}   %       for special symbols
\usepackage{enumerate}

\begin{document}
%==============================================================================
\title{Higher order net-proton number cumulants dependence on the centrality definition and other spurious effects}

\author{S.~Sombun$^3$, J.~Steinheimer$^1$, C.~Herold$^3$, A.~Limphirat$^3$, Y.~Yan$^3$ M.~Bleicher$^{1,2,4,5}$}

\address{$^1$ Frankfurt Institute for Advanced Studies, Ruth-Moufang-Str. 1, 60438 Frankfurt am Main, Germany}
\address{$^2$ Institut f\"ur Theoretische Physik, Goethe Universit\"at Frankfurt, Max-von-Laue-Strasse 1, D-60438 Frankfurt am Main, Germany}
\address{$^3$ School of Physics and Center of Excellence in High Energy Physics $\&$ Astrophysics, Suranaree University of Technology, Nakhon Ratchasima 30000, Thailand}
\address{$^4$ GSI Helmholtzzentrum f\"ur Schwerionenforschung GmbH, Planckstr. 1, 64291 Darmstadt , Germany}
\address{$^5$ John von Neumann-Institut f\"ur Computing, Forschungzentrum J\"ulich,
52425 J\"ulich, Germany}

\begin{abstract}
We study the dependence of the normalized moments of the net-proton multiplicity distributions on the definition of centrality in relativistic nuclear collisions at a beam energy of $\sqrt{s_{\mathrm{NN}}}= 7.7$ GeV. Using the UrQMD model as event generator we find that the centrality definition has a large effect on the extracted cumulant ratios. Furthermore we find that the finite efficiency for the determination of the centrality introduces an additional systematic uncertainty. Finally, we quantitatively investigate the effects of event-pile up and other possible spurious effects which may change the measured proton number. We find that pile-up alone is not sufficient to describe the data and show that a random double counting of events, adding significantly to the measured proton number, affects mainly the higher order cumulants in most central collisions. 
\end{abstract}

%\maketitle
\ioptwocol
\section{Introduction}

The collision of heavy ions allows to explore the properties of the strong interaction, called Quantum Chromodynamics (QCD) in the laboratory. Especially the phase structure of QCD is still under intense investigation, both experimentally and theoretically. A particularly interesting feature of this phase structure is the possible existence of a first order phase transition which ends in a critical endpoint. Recently, a focus of different experinmental programs was the measurement of the cumulants of net-proton number distributions, as they show significant deviations from the Poisson baseline, close to the phase transition \cite{Steinheimer:2012gc,Chomaz:2003dz,Randrup:2003mu,Sasaki:2007db} and the critical endpoint \cite{Stephanov:1998dy,Stephanov:2008qz,Koch:2008ia}.
In the last decade, tremendous progress has been made on the theoretical side by the calculation of (higher order) susceptibilities of baryon number, strangeness and other charges \cite{Gupta:2011wh,Luo:2011rg,Luo:2017faz,Herold:2016uvv,Zhou:2012ay,Wang:2012jr,Karsch:2011gg,Schaefer:2011ex,Chen:2011am,Fu:2009wy,Cheng:2008zh}. On the experimental side the measurements of these susceptibilities via event-by-event fluctuations has been pushed forward with experiments at RHIC \cite{Aggarwal:2010wy,Adamczyk:2013dal,Adamczyk:2014fia,Luo:2011ts,Adare:2015aqk} and at LHC \cite{Rustamov:2017lio,Abelev:2012pv}. In fact, first measurements of 6th order and even 8th order cumulants or moments of the distributions have been presented recently or are currently under investigation \cite{Chen:2016xyu,Friman:2011pf}. In spite of these progresses, a detailed understanding and interpretation of the measured moments is difficult due to a) uncertainties in the centrality determination which is crucial to avoid volume fluctuations, b) efficiency corrections, and c) transverse momentum ($p_T$) cuts that have to be extrapolated to low $p_T$ on an event-by-event basis.
Such experimental uncertainties can be, in part, studied by the use of transport models as event generators \cite{He:2017zpg}.
The present paper elucidates some of the above mentioned effects and their influence on the moments of the net-proton distributions by using the UrQMD model \cite{Bass:1998ca,Bleicher:1999xi}.

\subsection{The UrQMD model}
We will use the Ultra relativistic Quantum Molecular Dynamics (UrQMD) transport model, which is based on binary elastic and inelastic scattering of hadrons, including resonance excitations and decays as well as string dynamics and strangeness exchange reactions  \cite{Bass:1998ca,Bleicher:1999xi,Graef:2014mra}. The model is based on a geometrical interpretation of scattering cross sections which are taken, when available, from experimental data \cite{Olive:2016xmw} or model calculations, e.g. the additive quark model. In our investigations we use the UrQMD model in its cascade version, i.e. we do not include effects of long range potentials. Therefore, the model serves as a baseline event generator which includes only known vacuum physics. Since the cascade version of the model is not very computationally intensive we are able to accumulate sufficient statistics to study different 'non-physical' effects even on higher order cumulants of the net-proton number distributions. 

\section{Method}\label{method}

In the following we will present results of the sensitivities of cumulants of the multiplicity distribution of net-protons in high energy nuclear collisions on different 'non-physical' effects like centrality definition and acceptance and efficiency. The first four cumulants are defined as:
\begin{eqnarray}
C_1 &=& M  = \left\langle \mathrm{N} \right\rangle \\
C_2 &=&\sigma^2 = \left\langle (\delta \mathrm{N})^2 \right\rangle \\
C_3 &=& S \sigma^{3} =\left\langle (\delta \mathrm{N})^3 \right\rangle \\
C_4 &=& \kappa \sigma^{4} =\left\langle (\delta \mathrm{N})^4 \right\rangle - 3 \left\langle (\delta \mathrm{N})^2 \right\rangle^2 
\end{eqnarray}
 where $\delta \mathrm{N}= \mathrm{N}-\left\langle \mathrm{N} \right\rangle$ with N being the net-proton number in a given event and the brackets denoting an event average. Here $M$ is the Mean, $\sigma^2$ the variance, $S$ the Skewness and $\kappa$ the Kurtosis of the underlying multiplicity distribution.\\
 Usually one takes the following appropriate ratios of these cumulants:
 \begin{eqnarray}
C_2/C_1 &=& \sigma^2/M \\
C_3/C_2 &=&S \sigma\\
C_4/C_2 &=& \kappa \sigma^{2} 
\end{eqnarray}
For a Poisson distribution these ratios will always be equal to 1. These ratios also have the advantage that they cancel out the volume dependence of the cumulants, The dependence on fluctuations of the volume, however, is still present \cite{Skokov:2012ds,Begun:2016sop}. 

\begin{figure}[t]	%       -----------------------------------------
\includegraphics[width=0.5\textwidth]{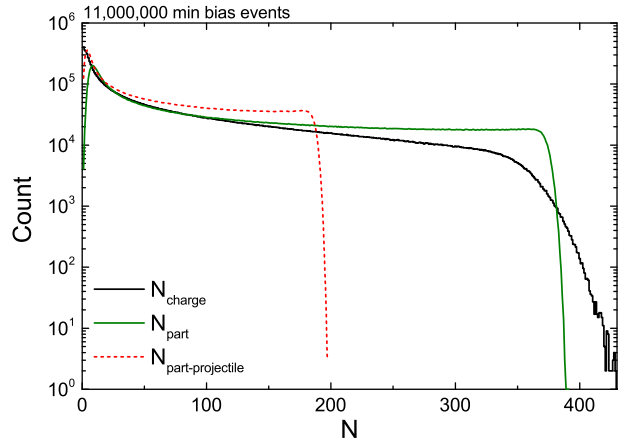}
\caption{[Color online] Distributions of $N_{charge}$, $N_{part}$ and $N_{\rm part-projectile}$ for minimum bias collisions of Au+Au nuclei at a beam energy of $\sqrt{s_{\mathrm{NN}}}= 7.7$ GeV. 
}\label{f1}
\end{figure}		%       ----------------------------------------- 

If one wants to compare experimental results on the cumulants of a nuclear collision to results from a model calculation, one faces several problems. For example, 
there are several ways to define centrality in the data and/or in the model, either by the impact parameter $b$, the number of participants $\mathrm{N}_{\rm part}$, or number of charged particles $\mathrm{N}_{\rm charge}$ in a given acceptance range. The impact parameter cannot be measured directly from experiment, and only average values of $b$ can be inferred from comparisons with models. Thus, for an event-by-event comparison between data and model, the impact parameter $b$ is not very useful.
That leaves $\mathrm{N}_{\rm part}$ and $\mathrm{N}_{\rm charge}$ as possible estimators for the centrality of a collision. Depending on the experimental specifications, either quantity is taken as a proxy of the volume of the system in a given event. Calculating the cumulants for a specific value of $\mathrm{N}_{\rm part}$ or $\mathrm{N}_{\rm charge}$ is then believed to minimize the fluctuations of the volume which can have an influence on the extracted cumulant ratios. Here it is important to note that the concept of a volume is ill-defined in a rapidly expanding system of small size, as it is produced in nuclear collisions. At no time during the evolution will there be a significantly large part of the system which can be characterized as being in global equilibrium with a given temperature and chemical potential, freezing out instantly. As we will see later, the centrality selection by $\mathrm{N}_{\rm part}$ and $\mathrm{N}_{\rm charge}$ can only be a rough approximation of the effective volume of the system at particle freeze out.\\

\begin{figure}[t]	%       -----------------------------------------
\includegraphics[width=0.5\textwidth]{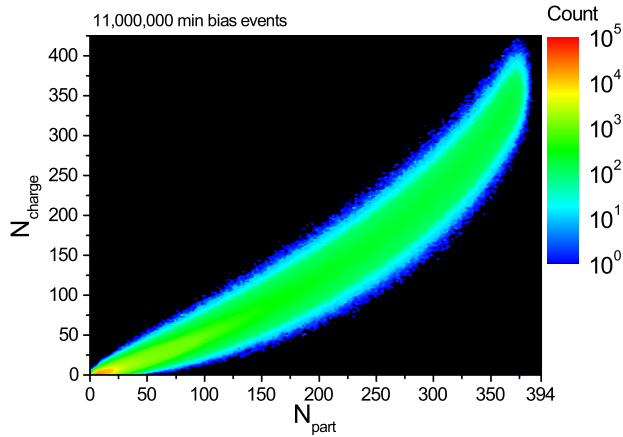}
\caption{[Color online] Distribution of $N_{\rm charge}$ as function of $N_{\rm part}$ for 11 million minimum bias events. 
}\label{f2}
\end{figure}		%       ----------------------------------------- 

To study the dependence of the net-proton number cumulants on the different centrality definitions we first have to specify these definitions. We will demonstrate the procedure for minimum bias Au+Au collisions at a beam energy of $\sqrt{s_{NN}}=7.7$~GeV. We define the following quantities:
\begin{itemize}
\item N$_{\rm charge}$: The number of all charged particles with $|\eta|\le 1$ and $0.15<p_T<2.0$~GeV minus the number of protons and anti-protons in this specific acceptance range.
\item N$_{\rm part}$: 394 minus all spectator protons and neutrons defined by $|y|>1.5$ and $p_T<0.3$~GeV.
\item N$_{\rm part-projectile}$: 197 minus all projectile spectator protons and neutrons defined by $y>1.5$ and $p_T<0.3$~GeV.
\end{itemize}
Note that in this study we define a spectator different from other transport model studies where a spectator is strictly a nucleon which has not undergone any scattering. In experiments such a definition is not measurable. Our definitions are therefore motivated by the experimental definition of a spectator. However, we have checked that indeed definition by number of scatterings and our rapidty and $p_T$ cut give similar results for the number of spectators.
For a total of 11 million minimum bias events the distributions for these quantities, calculated by UrQMD, is shown in figure \ref{f1}. The three different methods expectedly give different distributions. While the participiant distributions show a sharp cutoff at the maximum number of participants, the $N_{\mathrm{\rm charge}}$ distribution shows a much smoother drop. 

\begin{figure}[t]	%       -----------------------------------------
\includegraphics[width=0.5\textwidth]{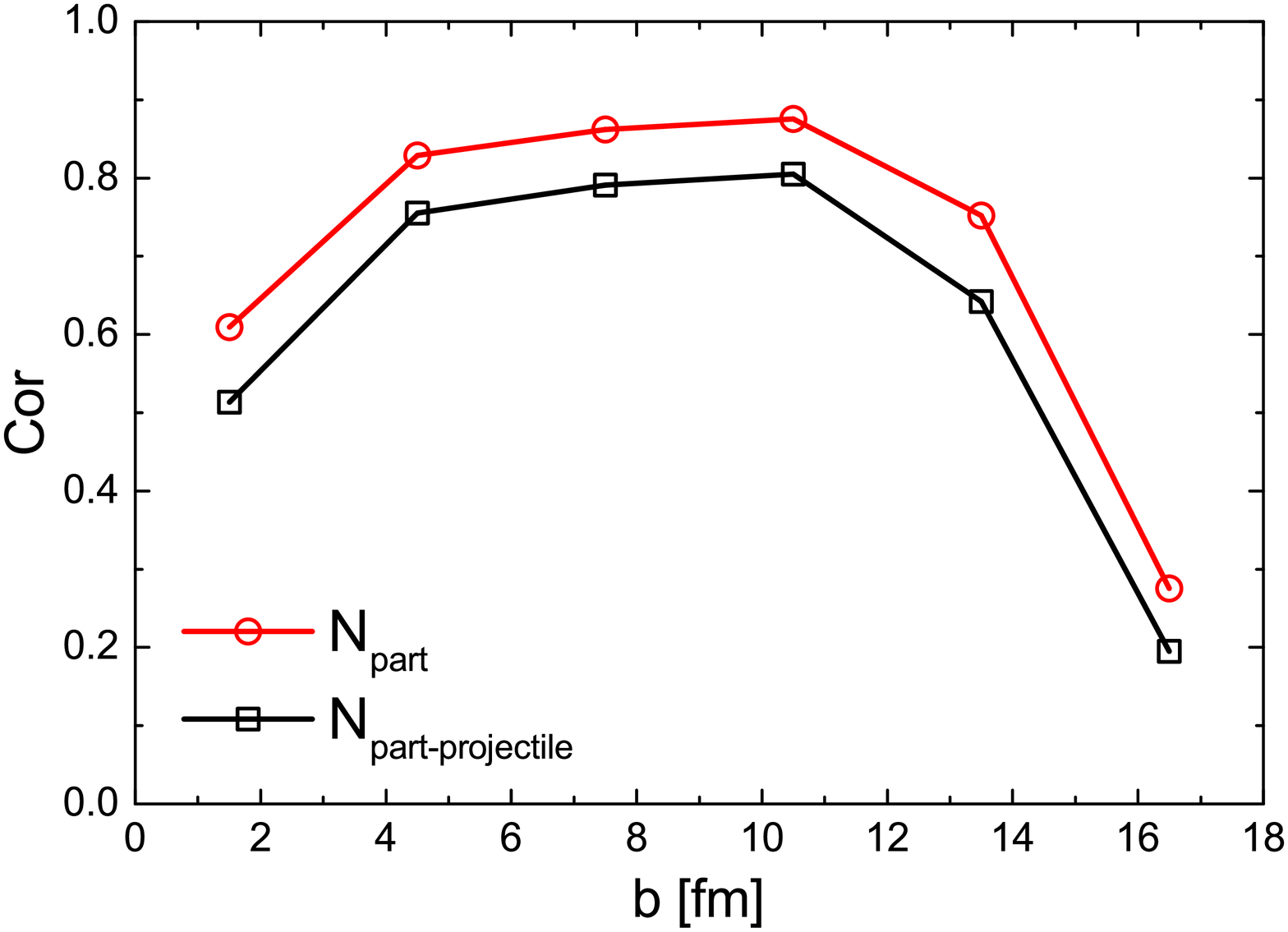}
\caption{[Color online] Correlation coefficient (see text) between $N_{\rm charge}$ and $N_{\rm part}$ as function of the impact parameter for minimum bias collisions of Au+Au nuclei at a beam energy of $\sqrt{s_{\mathrm{NN}}}= 7.7$ GeV. 
}\label{f3}
\end{figure}		%       ----------------------------------------- 

\begin{figure*}[t]	%       -----------------------------------------
\includegraphics[width=\textwidth]{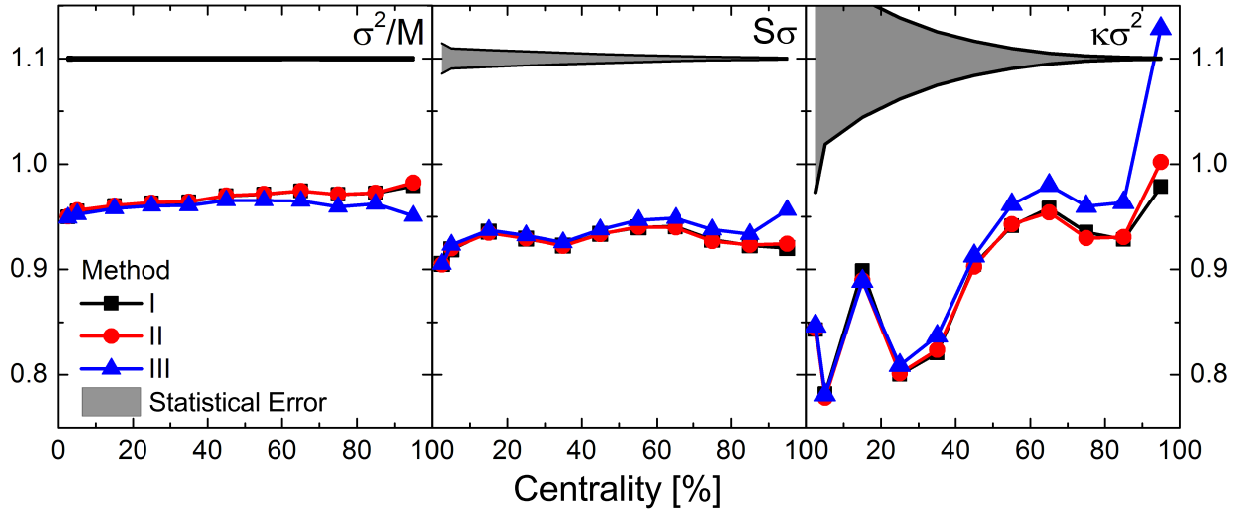}
\caption{[Color online] Variance, skewness and kurtosis of the net-proton number in different centrality bins calculated with the three different averaging methods described in the text. The grey band indicates the statistical error.
}\label{f5}
\end{figure*}		%       - ---------------------------------------- 

\section{Results}

First, we aim to understand the correlation between the number of charged particles produced in a single event and the number of participants in that same event. The distribution of the number of charged particles and number of total participants in an event is shown in figure \ref{f2}. There is an apparent correlation, however, the resulting distribution also shows a significant width in both $N_{\rm charge}$ and $N_{\rm part}$.

The correlation between the number of measured charged particles and participants can be quantified by the linear correlation coefficient $cor$. It is defined as:
\begin{equation}
cor=\frac{\sum_i (N_{\rm charge}^{i}- \left\langle N_{\rm charge}\right\rangle)(N_{\rm part}^{i}- \left\langle N_{\rm part}\right\rangle)}{\sigma_{\rm charge} \sigma_{\rm part}}~,
\end{equation}
where the sum always runs over all events in a given range of impact parameters $b$, with $\Delta b = 3$~fm.

The resulting correlation coefficient is shown in figure~\ref{f3}. Both the correlation for $N_{\rm part}$ and $N_{\rm part-projectile}$ with $N_{\rm charge},$ shows a similar strength and centrality dependence. While very peripheral and very central events show the weakest correlation between number of participants and number of charged particles, the mid-central events show the strongest correlations. This is understandable as for peripheral collisions the number of participants only changes a little, and is limited to $N_{part}>1$, while the number of charged particles varies more strongly. For the most central collisions the picture is reversed as the number of participants is limited and the number of charged particles fluctuates significantly. 

In the following, we will show how the net-proton number cumulants extracted from the model simulations depend on the different definitions of centrality.
For all following results we will use 11 million minimum bias events generated with the UrQMD transport model, used in cascade mode. Note that this amount of events is significantly larger than what is currently available for the STAR experiment, at the beam energy of $\sqrt{s_{\mathrm{NN}}}= 7.7$ GeV.  

We will also show results as function of centrality (in $[\%]$) which corresponds to the most central percentage of events (with a given centrality definition). For example, a centrality of 10$\%$ would correspond to the 10$\%$ of events with the largest number of $N_{\rm charge}$ (or $N_{\rm part}$).
As is done in experimental analyses, we will also show the resulting cumulant ratios averaged for centrality bins of 10$\%$ width. That means that in principle there are 3 methods to average the extracted cumulants (or moments) over a given centrality bin:
\begin{enumerate}[I]
\item First calculate the cumulants $C_n$ for a fixed $N_{\mathrm{part}}$ and then average the cumulants over all $N_{\mathrm{part}}$ in a given centrality bin. The ratios of cumulants are then taken as ratios of averages \cite{Luo:2017faz}.
\item Calculate the ratios of the cumulants for a given $N_{\mathrm{part}}$ and then average the ratios over the centrality bin.
\item Calculate the variance, skewness, and kurtosis ($\sigma$, S and $\kappa$) for a given $N_{\mathrm{part}}$ and average them over a given centrality bin. Then take the appropriate average ratios \cite{Luo:2013bmi} to obtain the cumulant ratios.
\end{enumerate}

Note that only for a sufficiently smooth dependence of the cumulants on centrality, all three methods should give similar results, while strong variations or a steep increase will lead to varying results of the three methods.

The statistical errors in our simulations are estimated according to the delta-theorem \cite{Luo:2011tp}.
For normally distributed variables the errors of the cumulant ratios then are:
\begin{equation}\label{delth}
error(C_r/C_2) \propto \sigma^{r-2}/\sqrt{n}~,
\end{equation}
where $n$ is the number of events and $\sigma^2= C_2$ the variance of the observable.

The resulting cumulant ratios of the net-proton number distributions are shown in figure \ref{f5}.

From our results we can say that (given sufficient statistics) all methods do agree well, except for the most peripheral bins. In these peripheral centrality bins the values of the cumulants appear to be changing rapidly with centrality and thus the average cumulant ratios depend on which method is used to calculate them. As a consequence we will apply method I for all the following studies.

\begin{figure}[t]	%       -----------------------------------------
\includegraphics[width=0.5\textwidth]{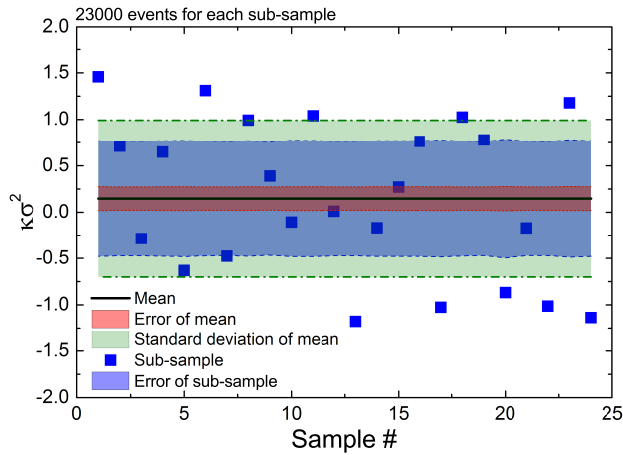}
\caption{[Color online] Kurtosis ($C_4/C_2$) for 24 different sub samples (blue squares) of the overall distribution which has a mean kurtosis value given by the black line. The red are indicates the error of the mean, according to the delta-theorem. The blue area is the error of each sub sample according to the delta-theorem. The green area indicates the actual standard deviation of the blue squares with respect to the mean. 
}\label{f4}
\end{figure}

Next, we want to find out whether the error estimates, using the delta theorem (\ref{delth}) are a good representation of the actual errors of the cumulants. To do so, we randomly subdivided our 0-5$\%$ total sample of 550000 events into 24 sets of sub-events. For each set of sub-events we then calculated the kurtosis of that sample, shown as blue squares in figure \ref{f4}. These values are now randomly distributed around the mean value of these sub-event sets shown as the black solid line. Furthermore, we calculated the error of the mean and sub-sample according to the delta theorem (red and blue bands) as well as the standard deviation of the mean (green shaded area). We find that the errors of the sub-samples are slightly smaller than the standard deviation of the mean value which means that the error estimation by the delta theorem is a little bit too small for the sample we have considered. However, the deviation is only small and we can say that the delta theorem gives a good approximation of the statistical error.

		%       ----------------------------------------- 

\begin{figure}[t]	%       -----------------------------------------
\includegraphics[width=0.5\textwidth]{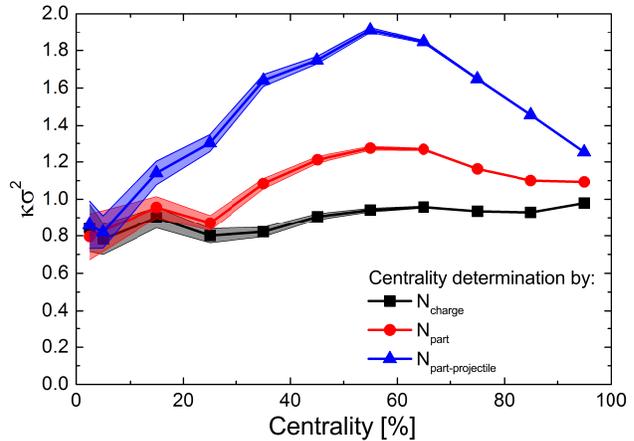}
\caption{[Color online] The net-proton kurtosis as function of centrality, which is defined by three different quantities, the number of charged particles, the number of participants or the number of participants in the projectile hemisphere. 
}\label{f8}
\end{figure}		%       ----------------------------------------- 

\subsection{Dependence on centrality definition}
First, we want to quantify the effect that the definition of centrality has on the observed cumulant ratios. As we have discussed earlier, different experiments may use different observables to define centrality due to their varying acceptance and detector components. Figure \ref{f8} shows the scaled kurtosis of protons at mid-rapidity $|y|<0.5$, as function of the centrality, using method I to average over a centrality bin, for collisions of Au nuclei at a center of mass  energy of $\sqrt{s_{\mathrm{NN}}}=7.7$ GeV. Here we assume a limited acceptance in transverse momentum $0.4<p_T<0.8$ GeV. The centrality is defined by three different methods explained in section \ref{method}. For now we assumed a perfect detection efficiency and acceptance for all particles used to determine the centrality. We observe an interesting dependence on the centrality definition. For the most central events, the value of the kurtosis does not depend on the choice of centrality selection, even though the different definitions are only weakly correlated, as shown in figure \ref{f3}. This indicates that, even though the measures of centrality are weakly correlated, the effective volume is essentially fixed and 'volume  fluctuations' do not contribute strongly to the observed cumulants. On the other hand, as we increase the centrality, the difference between the definitions become large. In particular the case where only the projectile participants are fixed shows a prominent increase of the cumulants for 40-60$\%$ most central collisions. This is understandable as the number of target participants is allowed to fluctuate, thus increasing the measured cumulants significantly. The smallest dependence of the cumulant ratios is observed when the number of charged particles is used for the centrality definition. This may be due to the fact that the charged particles are also measured around mid-rapidity thus their multiplicity is stronger correlated with the mid-rapidity 'volume' and therefore with the number of protons at mid-rapidity.

\begin{figure}[t]	%       -----------------------------------------
\includegraphics[width=0.5\textwidth]{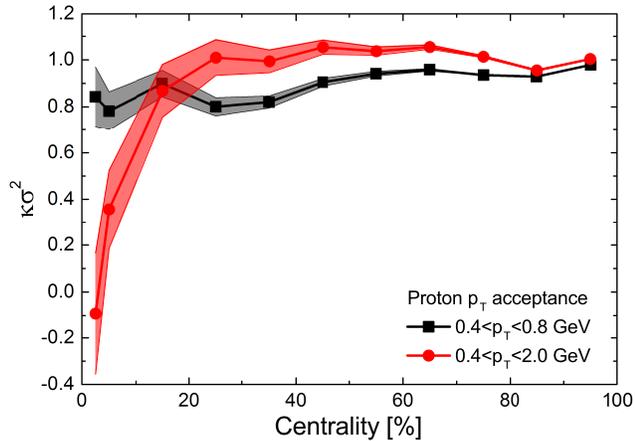}
\caption{[Color online] The net-proton kurtosis as function of centrality (defined by $N_{\rm charge}$) for two different $p_{T}$ acceptance selections. 
}\label{f9}
\end{figure}		%       ----------------------------------------- 

\subsection{Effects of acceptance}

Next, we investigate the acceptance dependence of the net-proton number kurtosis for the same system as discussed above. This time however, we only use charged particles $N_{\mathrm{charge}}$ for the centrality selection. Figure \ref{f9} shows the results for two different cuts in transverse momentum for the protons and anti-protons used to calculate the kurtosis. A significant effect is observed for the most central collisions, where the value of the kurtosis is significantly reduced for the larger $p_T$ acceptance. This can be understood as an result of the global conservation of baryons which will reduce the kurtosis. On the other hand, for mid-central collisions the scaled kurtosis is actually larger for the case of an increased acceptance which again indicates that so-called volume fluctuations dominate the extracted kurtosis and the effects of baryon conservation are not visible in the results.

\begin{figure}[t]	%       -----------------------------------------
\includegraphics[width=0.5\textwidth]{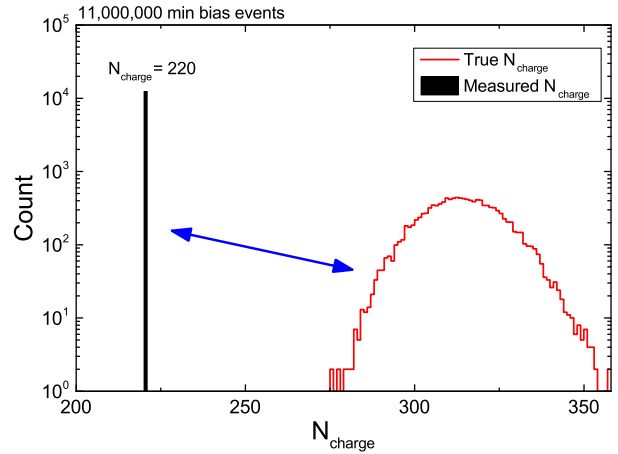}
\caption{[Color online] Assuming a 70$\%$ efficiency for charged particles we show the original $N_{\rm charge}$ distribution of a measured fixed $N_{\rm charge}$ bin. 
}\label{f10}
\end{figure}		%       ----------------------------------------- 

\begin{figure}[b]	%       -----------------------------------------
\includegraphics[width=0.5\textwidth]{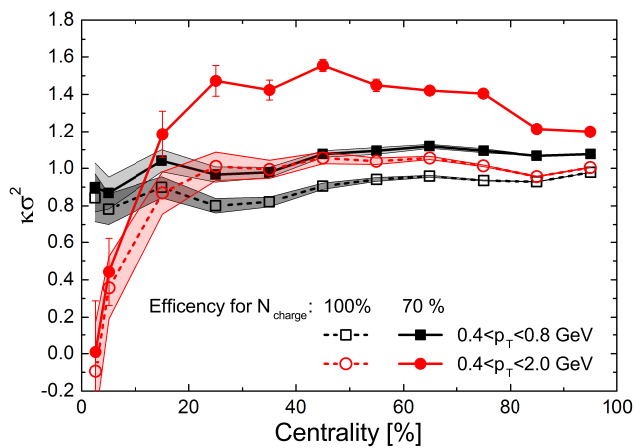}
\caption{[Color online] Kurtosis of the net-proton number in two different $p_{T}$ acceptance bins (black squares and red circles). We compare results with full efficiency for $N_{\rm charge}$ (open symbols) with results where we assume a 70$\%$ efficiency for $N_{\rm charge}$ (full symbols).
}\label{f11}
\end{figure}		%       ----------------------------------------- 

\subsection{Effects of efficiency}

\begin{figure*}[t]	%       -----------------------------------------
\includegraphics[width=\textwidth]{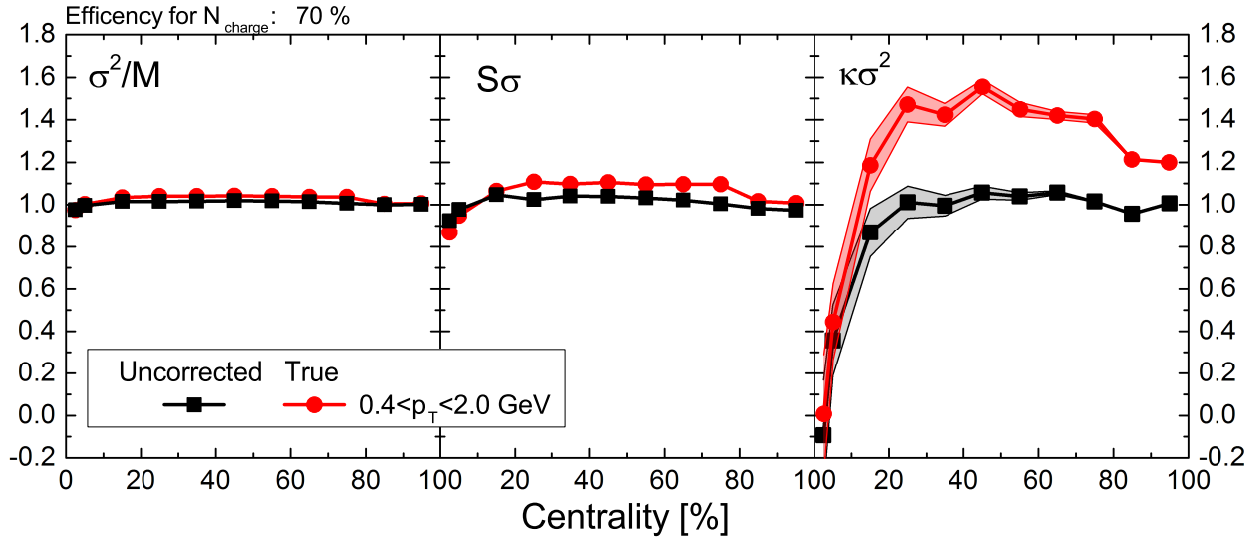}
\caption{[Color online] Scaled Variance, Skewness and Kurtosis of the net-proton number in two different $p_{T}$ acceptance bins (open and full symbols) for a 70$\%$ $N_{\rm charge}$ efficiency. We compare results with full efficiency for protons (red circles) with results where we assume a constant efficiency for protons (black squares). 
}\label{f12}
\end{figure*}		%       ----------------------------------------- 

An important shortcoming of experimental measurements of particle number fluctuations is the fact that detectors will never perform a perfect measurement. That means that in a single event there is only a certain probability that a produced particle is actually registered by the detector. This probability is called efficiency (see \cite{Abelev:2008ab} for the STAR efficiency). It is clear that if the efficient of a detector is less than 100$\%$ this will have an effect on all the measured particle number cumulants. While the correction for the first order cumulant, the mean particle number, can usually be done reasonably well, the corrections for higher order cumulants can become very complicated. Analytical formulas to correct the cumulants for a constant binomial efficiency have been derived in \cite{Bzdak:2013pha}, but in the case of a momentum dependent efficiency, these corrections cannot be done by simple analytic formulae \cite{Bzdak:2016qdc}.

The imperfect efficiency of the detectors has two distinct consequences. One is the one discussed above where the protons used to calculate the proton number in a given event cannot be measured exactly and thus the net-proton number cumulants have to be corrected. The second effect is more indirect as also the particles used to determine the centrality of an event cannot be measured exactly for each event, leading to a centrality determination efficiency. This effect is more indirect as it does not enter in the calculation of the cumulants but the determination of the centrality bin, thus it mixes events of different centrality and acts similar to volume fluctuations.

To demonstrate the effect of this centrality determination efficiency, we show in Fig. \ref{f10} the original distribution of the number of charged particles in all events that have been identified (after efficiency loss) to contain 220 charged particles. For simplicity we have assumed a binomial efficiency of 70$\%$ for charged particles, while in reality the efficiency may depend on momentum or event multiplicity, making the effect even more difficult to disentangle. But even for our simple binomial efficiency we see that the events which are identified as having $N_{\mathrm{charge}}=220$ actually originate from events with a very broad distribution of $N_{\mathrm{charge}}$.

The effect of the additional 'fake' volume fluctuations on the net proton number kurtosis is shown in Fig. \ref{f11} for two different $p_T$ acceptance cuts. In general, the smaller efficiency leads to an increase of the kurtosis. This effect, however, is largest for intermediate centralities, as is always the case for volume fluctuations. The more central events are less affected probably due to the fact that now the most central events correspond to events with large efficiency as well as large particle number, thus decreasing the effect of the efficiency fluctuations. One should note that in our case the efficiency for detection of protons is still considered 100$\%$. In our study we can only consider a scenario where the proton efficiency, in an event, is independent of the charged particle efficiency, thus there would be no correlation between events with a large number of charged particles and a good net-proton efficiency. However, it has to be considered that in a realistic detector setup the efficiency for detecting $N_{\mathrm{charge}}$ and the protons can be correlated and thus very central events could correspond to events with larger proton efficiency which would have a significant impact on the measured cumulants.

Taking into account also an imperfect (constant) $75\%$ binomial efficiency for protons, we show the results for the scaled variance, the skewness, and the kurtosis of the net-proton number in Fig. \ref{f12}. Here we especially compare the 'true' cumulants (red circle symbols) with perfect proton efficiency but 70$\%$ efficiency for $N_{\mathrm{charge}}$, and the 'uncorrected' cumulants where we assumed also the $75\%$ binomial efficiency for protons. As expected, the efficiency 'uncorrected' cumulants are smaller than the 'true' ones, which is also observed for the experimental data.
The decrease of the cumulants due to efficiency is, however, smallest for the most central events which is an opposite behavior to what is observed in the analysis of the STAR experiment where central events show the largest efficiency corrections \cite{Luo:2017faz}.

\subsection{Rapidity dependence}
\begin{figure}[t]	%       -----------------------------------------
\includegraphics[width=0.5\textwidth]{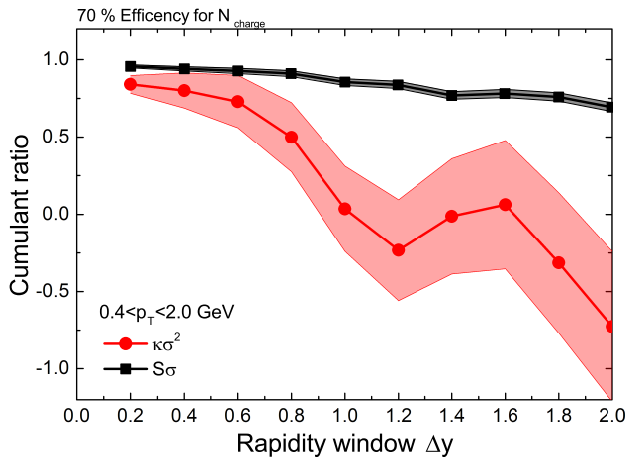}
\caption{[Color online] Kurtosis and skewness as function of the rapidity window $\Delta y$, around mid-rapidity. We assume again a 70$\%$ efficiency for charged particles and show only results for the larger $p_T$ acceptance.
}\label{f16}
\end{figure}

In the following, we will discuss the dependence of the cumulant ratios on the size of the rapidity acceptance $\Delta y$. In figure \ref{f16} we show the dependence of the kurtosis and skewness on the size of the rapidity window, $\Delta y$, in which the protons are measured. As one expects, for an increasing window the effects of baryon number conservation increase and therefore the cumulant ratios decrease. This effect is stronger for the kurtosis as it is for the skewness. For very small rapidity windows the cumulants should resemble those of a Poisson distribution, i.e. they should converge towards 1, which is also observed.

		%       ----------------------------------------- 

\subsection{Pile-up}
\begin{figure}[t]	%       -----------------------------------------
\includegraphics[width=0.5\textwidth]{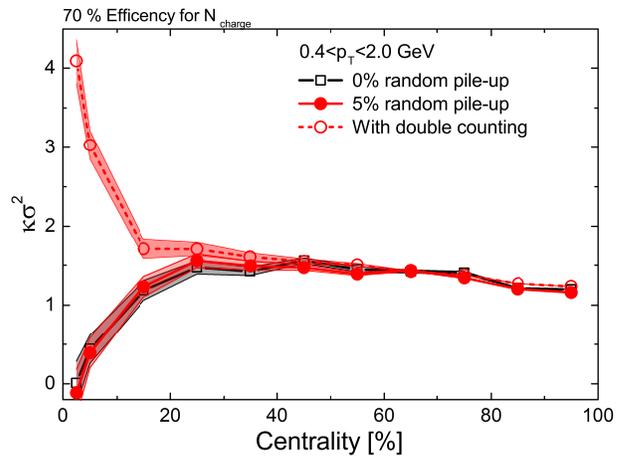}
\caption{[Color online] Kurtosis of the net proton number as function of centrality, defined by $N_{\rm charge}$. The efficiency for $N_{\rm charge}$ is $70\%$. We compare two cases, one with five percent pileup probability (full symbols) and one with an additional double counting of events.  
}\label{f15}
\end{figure}

The last effect we will discuss is the so called pile-up effect. Usually, pile-up refers to events which occur in such short succession that the detector is not able to identify them as two separate events but records them as a single event. The probability for such a pile-up event to occur is usually considered very small. However, it has been discussed in recent publications \cite{Garg:2017agr} that this effect may play a role in the measured cumulants. We can easily study the effect of pile-up in our model. To do so we randomly sort our events and, going through them in random order, assign a certain probability that two consecutive events will be merged to a single one. After that we will apply again the binomial efficiency for $N_{charge}$ and calculate the cumulants as done above. The results for the kurtosis for a pile-up probability of $5\%$ are shown in Fig. \ref{f15} for a single $p_T$ bin. We observe no change of the kurtosis. 

To investigate what kind of experimental artifacts would be required in order to explain the experimentally measured large kurtosis of the recent STAR results we allow for an additional, artificial increase of the proton number by randomly double counting events. The STAR \cite{Luo:2017faz} results show a strong increase of the kurtosis and a decrease of the skewness for most central collisions, if a larger acceptance in transverse momentum is taken into account. 

\begin{figure}[t]	%       -----------------------------------------
\includegraphics[width=0.5\textwidth]{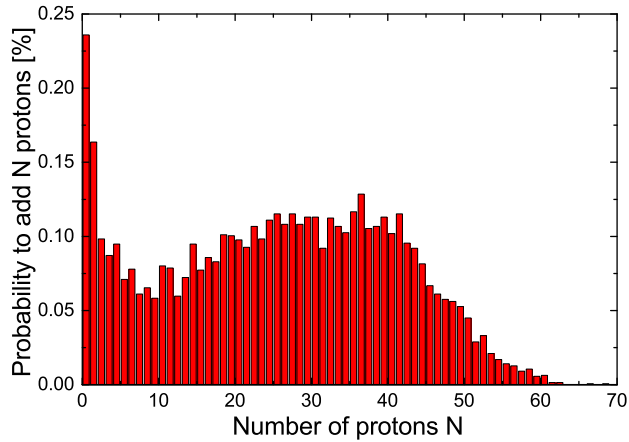}
\caption{[Color online] Probability distribution to find N additional protons in a central event. This probability distribution generates an increase of the kurtosis as shown in figure~\ref{f15}.  
}\label{f17}
\end{figure}

The probability distribution for adding N protons in a central events is shown in figure \ref{f17}.

One can see that the additional number of protons added in an event has to be on the order of 25, which is the mean of the surplus distribution. 

The large effect for the central events can be understood as a result of the 'lifting' of the global conservation of the net-baryon number due to the artificial increase in proton number. Since now the most central events can be made up by two full events, the total baryon number is now much less constrained which appears to be the main influence on the kurtosis of central events. However, such a large probability seems actually unreasonable only from pile-up events.
It is important to note that while randomly adding additional protons does give the desired results, an increase in the kurtosis, it does not decrease the skewness which would be in contrast to the STAR data. 
On the other hand there may be different effects which act similarly as pile-up, e.g. interactions between the beam and gas, misidentifications and others. Such experimental effects would also effectively remove the constraints of baryon number conservations and drastically change the higher order cumulants.  

		%       ----------------------------------------- 

		%       ----------------------------------------- 

\section{Conclusion}
We have explored different commonly used methods to define the centrality of the reaction. We found that the correlation between the $N_{\rm charge}$ estimation of the centrality and number of participant based estimators is by itself strongly centrality dependent and falls off to values around 0.5 for central and peripheral collisions. This indicates that the centrality determination method leads to a different mixing of different (true) centrality classes based on the used centrality estimator, leading to a systematic deviation of the moments calculated using the different estimators. This result is understood as for both the most central and most peripheral collisions, either the number of participants or number of charged particles is constrained while the other quantity fluctuates randomly.

Next we have explored the effect of two different transverse momentum ranges for the proton acceptance, here we found that for the top centrality classes a sizable difference between the extracted moments for the two ranges emerges, due to the importance of conservation laws. This difference gets small towards more peripheral collisions, where volume fluctuations dominate, however imposes a systematic uncertainty of the order of 10-20\%. 

Then the effect of the detector efficiency on the centrality determination using charge particle was explored. Also here, we observed systematic uncertainties of the order of 10-20\%, but also a stronger deviation in case of the larger $p_T$ acceptance window of up to 50\%, again due to conservation laws. We showed that the effect of finite efficiency on the centrality selection appears smaller for the most central events. This is because by selecting events with large $N_{\rm charge}$ one is biased towards events with a large (event-)efficiency in $N_{\rm charge}$, thus decreasing the effect of additional volume fluctuations. The strong effect for the larger acceptance window is a result of the global baryon number conservation. For a larger acceptance the contribution from the global conservation of the baryon charge becomes dominant, at least for the most central bins, and decreases the measured cumulants significantly.

The effect of conservation remains dominant even if a realistic (binomial) efficiency was taken into account. Our results therefore are in contrast to STAR data which show no effect of the global conservation laws on the measured proton number cumulants.
In the current study we could not investigate a scenario where the efficiency for measuring $N_{\rm charge}$ is correlated with the efficiency for measuring the number of protons in a given event. Such a correlation, possibly existent in experiment, would increase the extracted cumulants.

Finally we explored the effect of pile-up events. Here we observed that even a large pile-up probability will not result in a considerable modification of the measured cumulants. Only when we artificially double count events with a large proton number, i.e. we add by hand a large number of protons in a few events, the fourth order cumulant is significantly increased. 

In conclusion, we have addressed various sources of uncertainty for the estimation of higher moments (and their ratios) of conserved charge fluctuations. We found that the currently employed methods to obtain these moments from experimental data
may still suffer systematic uncertainties on the order of 20\%. In stark contrast to the STAR experiment, we always observe a very strong effect of global baryon number conservation for the most central events.

\section{Acknowledgments}
The authors thank V. Koch for helpful discussions.
The computational resources were provided by the LOEWE Frankfurt Center for Scientific Computing (LOEWE-CSC). This work was supported by the German Academic Exchange Service (DAAD), HIC for FAIR and the Thailand Research Fund (TRF). SS and AL acknowledge support from TRF-RGJ (PHD/0185/2558).   

%%%%%%%%%%%%%%%%%%%%%%%%%%%%%%%%%%%%%%%%%%%%%%%%%%%%%%%%%%%%%%%%%%%%%%%%%%%%%%%
%%%%%%%%%%%%%%%%%%%%%%%%%%%%%%%%%%%%%%%%%%%%%%%%%%%%%%%%%%%%%%%%%%%%%%%%%%%%%%% 
\end{document}